\documentclass[aps,prb,twocolumn,superscriptaddress]{revtex4}
\makeatletter

\newcommand{\Rmnum}[1]{\expandafter\@slowromancap\romannumeral #1@}
\makeatother

\usepackage{eurosym}
\usepackage{amsfonts}
\usepackage{amssymb}
\usepackage{amsmath}
\usepackage{graphicx}
\usepackage{color}
\begin{document}

\title{Edge-controlled half-metallic ferromagnetism and direct gap in ZrS$_{2}$ nanoribbons}

\author{H. Y. Lv}

\affiliation{Key Laboratory of Materials Physics, Institute of Solid State Physics, Chinese Academy of Sciences, Hefei 230031, People's Republic of China}

\author{W. J. Lu}
\email[Corresponding author: ]{wjlu@issp.ac.cn}
\affiliation{Key Laboratory of Materials Physics, Institute of Solid State Physics, Chinese Academy of Sciences, Hefei 230031, People's Republic of China}

\author{J. Y. Li}

\affiliation{Key Laboratory of Materials Physics, Institute of Solid State Physics, Chinese Academy of Sciences, Hefei 230031, People's Republic of China}

\author{R. C. Xiao}

\affiliation{Key Laboratory of Materials Physics, Institute of Solid State Physics, Chinese Academy of Sciences, Hefei 230031, People's Republic of China}

\author{M. J. Wei}

\affiliation{Key Laboratory of Materials Physics, Institute of Solid State Physics, Chinese Academy of Sciences, Hefei 230031, People's Republic of China}

\author{P. Tong}

\affiliation{Key Laboratory of Materials Physics, Institute of Solid State Physics, Chinese Academy of Sciences, Hefei 230031, People's Republic of China}

\author{X. B. Zhu}

\affiliation{Key Laboratory of Materials Physics, Institute of Solid State Physics, Chinese Academy of Sciences, Hefei 230031, People's Republic of China}

\author{Y. P. Sun}
\email[Corresponding author: ]{ypsun@issp.ac.cn}
\affiliation{Key Laboratory of Materials Physics, Institute of Solid State Physics, Chinese Academy of Sciences, Hefei 230031, People's Republic of China}
\affiliation{High Magnetic Field Laboratory, Chinese Academy of Sciences, Hefei 230031, People's Republic of China}
\affiliation{Collaborative Innovation Center of Advanced Microstructures, Nanjing University, Nanjing, 210093, People's Republic of China}

\makeatletter

%%%%%%%%%%%%%%%%%%%%%%%%%%%%%% LyX specific LaTeX commands.

\begin{abstract}
The electronic and magnetic properties of ZrS$_{2}$ nanoribbons (NRs) are investigated based on the first-principles calculations. It is found that the ZrS$_{2}$ NRs with armchair edges are all indirect-band-gap semiconductors without magnetism and the band-gap exhibits odd-even oscillation behavior with the increase of the ribbon width. For the NRs with zigzag edges, those with both edges S-terminated are nonmagnetic direct-band-gap semiconductors and the gap decreases monotonically as a function of the ribbon width. However, the NRs with one edge S-terminated and the other edge Zr-terminated are ferromagnetic half-metal, while those with both edges Zr-terminated tend to be ferromagnetic half-metal when the width $N\geq9$. The magnetism of both systems mainly  originates from the unsaturated edge Zr atoms. Depending on the different edge configurations and ribbon widths, the ZrS$_{2}$ NRs exhibit versatile electronic and magnetic properties, making them promising candidates for the applications of electronics and spintronics.

\end{abstract}
\maketitle

The electronic and magnetic properties of nanoscale materials have been the subject of extensive research due to their potential applications in electronics and spintronics. Carbon-based systems, graphene for example, are among the mostly studied low-dimensional materials. Graphene has a high mobility at room temperature, making it a promising candidate for the future information technology. However, the lack of intrinsic band gap has limited its practical applications. Cutting the two-dimensional (2D) graphene into one-dimensional (1D) nanoribbon (NR) can open the band gap in graphene, which was firstly predicted theoretically\cite{PRL-Louie} and then verified by the experiments.\cite{Science-Dai} Because of the additional edge states unique to the dimensionality, 1D NRs could exhibit rich properties, which can be further tuned by modifying their edges. It was found that the narrow zigzag graphene NR is semiconducting with the two edges antiferromagnetically coupled to each other.\cite{PRL-Louie} In addition, the graphene NR can be converted into half-metal in different ways, such as by applying a homogeneous electrical field\cite{Nature-Louie} or by chemical decoration at the edges.\cite{NanoLett-Hod,JACS-Kan}

Besides graphene, low-dimensional transition-metal dichalcogenides TMDs (with the formula MX$_2$, M=transition metal, X=S, Se, or Te) are another important materials that received considerable attention. Different from graphene, single layer MX$_2$ has three atomic layers, with an M-layer sandwiched between two-X layers. As a typical representative of TMDs, single layer MoS$_2$ is a direct-band-gap semiconductor.\cite{PRL-Mak,PRB-Kuc} Intrinsic 2D MoS$_2$ monolayer is nonmagnetic. However, distinct electronic and magnetic properties were reported for 1D MoS$_2$ NRs, that is, armchair MoS$_2$ NRs are nonmagnetic semiconductors, while the zigzag NRs have ferromagnetic (FM) and metallic ground states.\cite{JACS-Chen,Nanotech-Terrones} The WS$_2$ NRs exhibit similar properties as to MoS$_2$ NRs.\cite{JAP-Zhang} The zigzag-edge related ferromagnetism was then observed experimentally in MoS$_2$\cite{APL-Tongay,NRL-Xue,NRL-Gao,Nanoscale-Xue} and WS$_2$\cite{APL-Li} nanosheets. Furthermore, the electronic and magnetic properties of MoS$_2$ NRs can be modified by edge passivation\cite{JPCC-Zhang} or by applying external strain and/or electric field.\cite{JPCC-Pan,JPCL-Kou,ACSNano-Sanvito,JAP-Hu} It was theoretically predicted that by applying a transverse electrical field, the insulator-metal transition occurs and ferromagnetism emerges beyond a critical electrical field in the armchair MoS$_2$ NRs.\cite{ACSNano-Sanvito}

Both MoS$_2$ and WS$_2$ crystallize in the honeycomb (H) structure. ZrS$_2$ is another kind of TMDs, which crystallizes in the centered honeycomb (T) structure. Different from MoS$_2$ monolayer, single layer ZrS$_2$ is an indirect-band-gap semiconductor.\cite{RSCAdv-Li} In experiment, 1D ZrS$_2$ NRs have been synthesized by the process of chemical vapor transport and vacuum pyrolysis.\cite{CC-Zhang,Advmat-Li} Then what different properties the 1D NRs may present? In the present work, we investigate the electronic and magnetic properties of a series of ZrS$_2$ NRs. The results show that depending on  the different edge configurations, the ZrS$_2$ NRs can be indirect-band-gap semiconductor, direct-band-gap semiconductor, antiferromagnetic (AFM) metal, or FM half-metal, exhibiting versatile electronic and magnetic properties.

\begin{figure}[htpb]
\includegraphics[width=0.9\columnwidth]{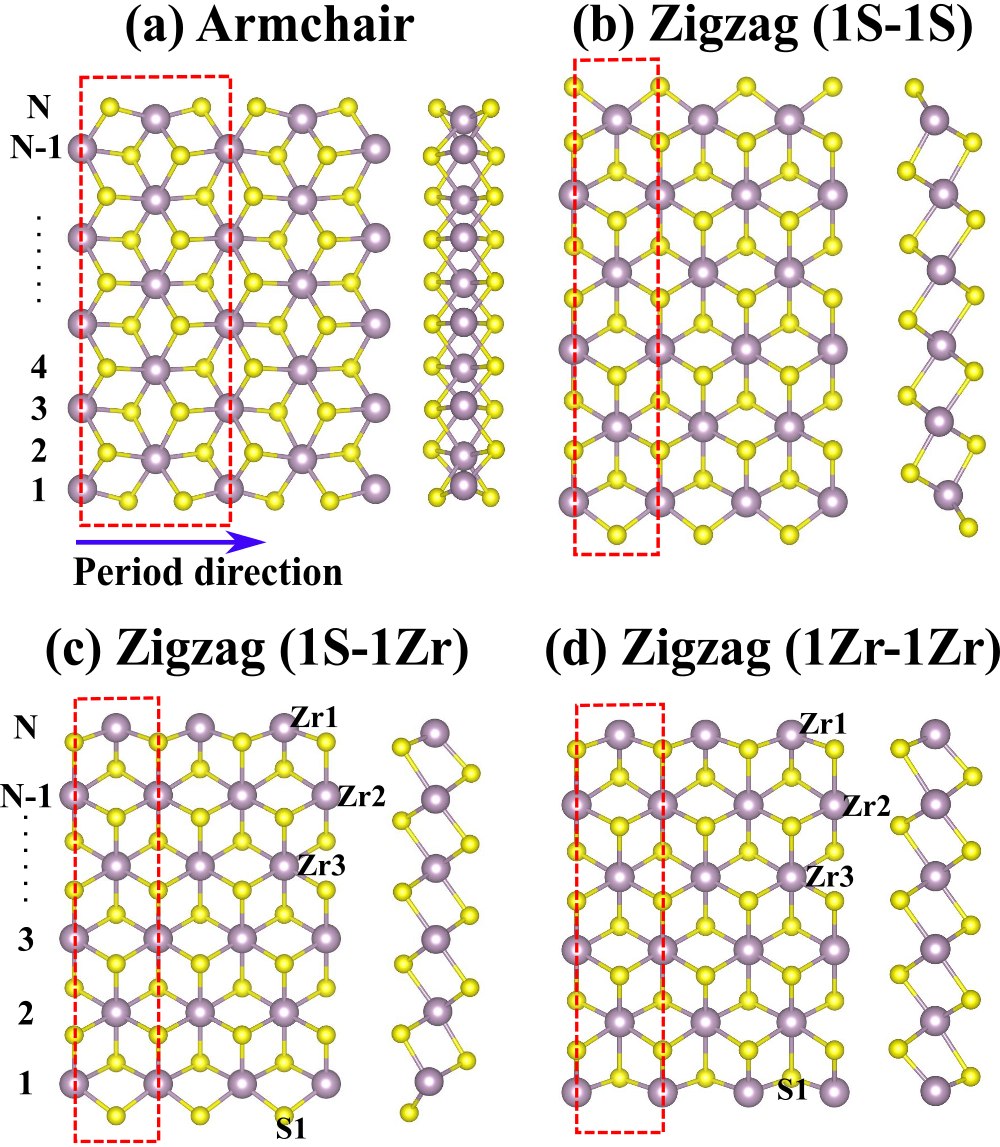}\caption{\label{fig1-structures}Top and side views of the four different configurations of ZrS$_2$ NRs: (a) armchair, (b) 1S-1S zigzag, (c) 1S-1Zr zigzag, and (d) 1Zr-1Zr zigzag NRs. The red dashed line indicates the unit cell. }
\end{figure}

Our calculations were performed within the framework of the density functional theory (DFT), \cite{PR-DFT} as implemented in the QUANTUM ESPRESSO code.\cite{QE} The exchange correlation energy was in the form of Perdew-Burke-Ernzerhof (PBE)\cite{PBE} with generalized gradient approximation (GGA). The Brillouin zones were sampled with $12\times1\times1$ and $20\times1\times1$ Monkhorst-Pack k meshes for the armchair and zigzag NRs, respectively. The cutoff energy for the plane-wave expansion was set to be 40 Ry. The distance between each NR and its periodic image is set to be larger than 15 {\AA} so that they can be treated as independent entities.

The ZrS$_2$ NRs can be obtained by cutting the monolayer, and there are mainly two kinds of ZrS$_2$ NRs, i.e., armchair and zigzag NRs, according to the cutting directions with respect to the monolayer. As for the zigzag ZrS$_2$ NRs, the edges can be terminated by Zr or S atoms, thus three cases exist. We denote the case that two edges are both S atoms terminated as 1S-1S zigzag NR, and the other two cases are 1Zr-1S (one edge is Zr and the other edge is S atoms) and 1Zr-1Zr (both edges are Zr atoms) zigzag NRs. The relaxed structures of these four kinds of ZrS$_2$ NRs are demonstrated in Fig. 1 and the left and right panels are top and side views, respectively. The width of the NR is defined according to the number of the Zr atoms across the ribbon width, as shown in Figs. 1(a) and (c). Compared with the initial structures, the fully relaxed structures of the NRs change a little, mainly coming from the edge atoms, that is, the edge S atoms tend to leave away from the inner atoms, while the Zr atoms prefer to approach the inner side.

\begin{figure}[htpb]
\includegraphics[width=0.8\columnwidth]{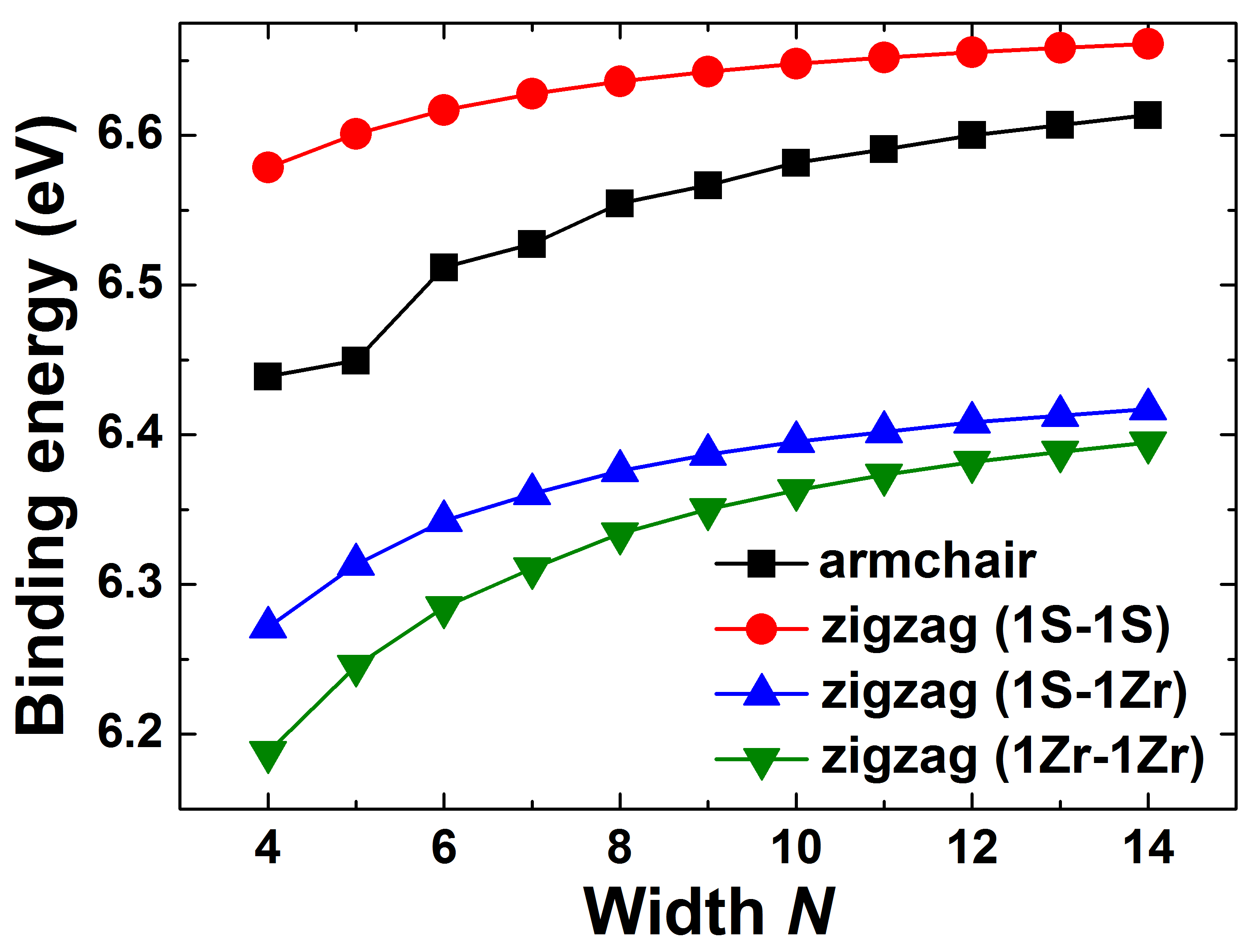}\caption{\label{fig2-Binding energy}Binding energies of the four kinds of ZrS$_2$ NRs as a function of ribbon width.}
\end{figure}

To investigate the stability of the ZrS$_2$ NRs, we calculated the binding energy $E_b$, which is defined as $E_b$=(E$_{Zr_nS_m}-nE_{Zr}-mE_S$)/$(n+m)$, where $E$(Zr$_n$S$_m$), $E$(Zr), and $E$(S) are the total energies of the ZrS$_2$ NRs, the Zr and S atoms, respectively. The larger the binding energy, the more stable the corresponding structure. The NRs with the width $N=4-14$ are considered in this work. The calculated binding energies of the four kinds of ZrS$_2$ NRs are in the range of 6.19$-$6.66 eV per atom, indicating that all the investigated NRs are very stable. Moreover, as shown in Fig. 2, the binding energies of the 1S-1S zigzag NRs are the highest among the four kinds of NRs, indicating that this type of NR is the most stable one, followed by the armchair NRs, and the 1S-1Zr and 1Zr-1Zr zigzag NRs are relatively less stable. For each kind of NRs, the system becomes more and more stable as the ribbon width increases.

\begin{figure*}[htpb]
\includegraphics[width=1.5\columnwidth]{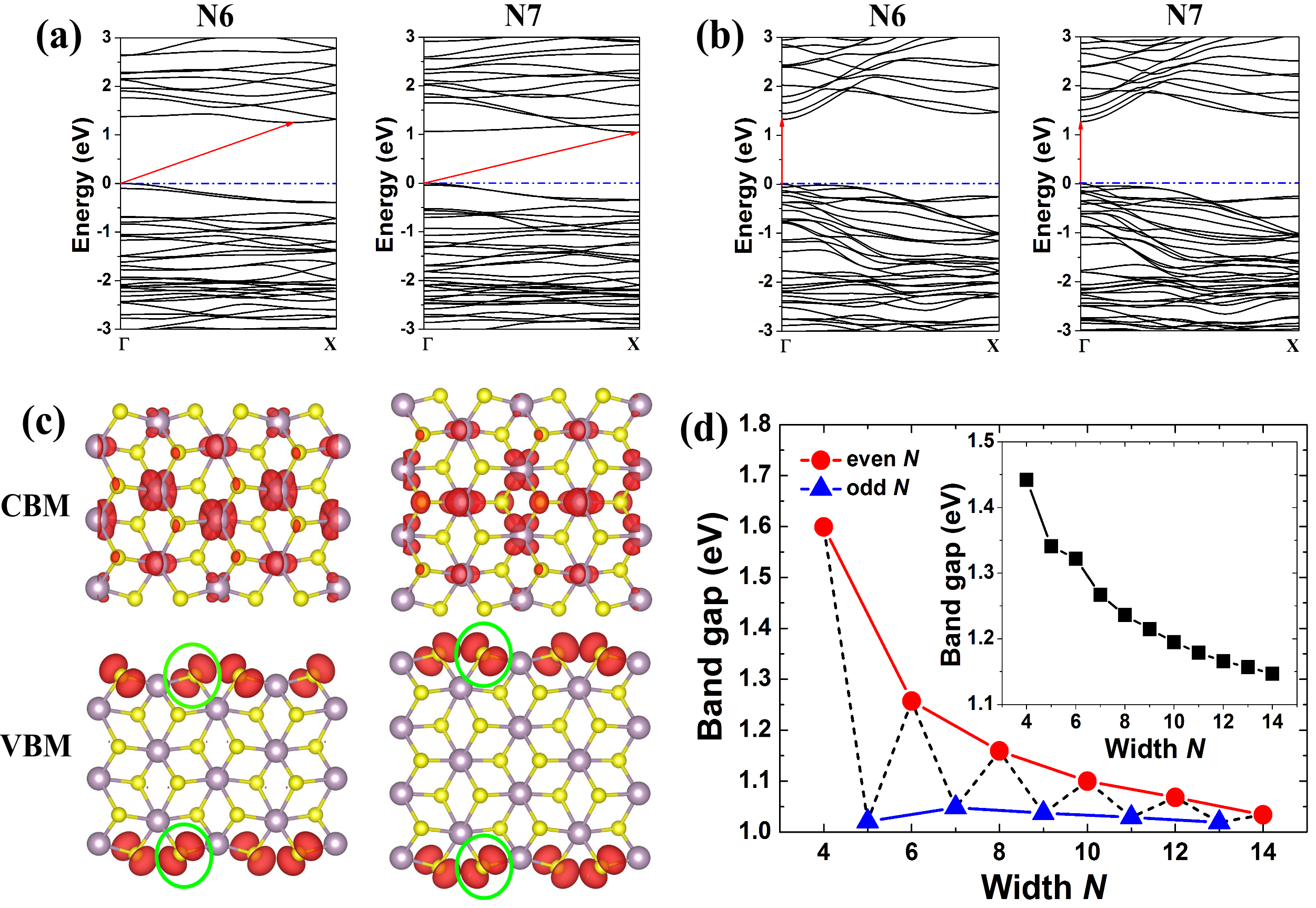}\caption{\label{fig3-Band gap}Band structures of the two kinds of semiconducting ZrS$_2$ NRs: (a) armchair and (b) 1S-1S zigzag NRs with width $N=6$ and 7. (c) The band-decomposed charge densities for the conduction band minimum (CBM) and valence band maximum (VBM) of armchair NRs with width $N=6$ (left panel) and $N=7$ (right panel). (d) Width-dependent band gap of the armchair ZrS$_2$ NR. The inset is the result for 1S-1S zigzag NRs.}
\end{figure*}

To check the possibility of magnetism in ZrS$_2$ NRs, both the spin-unpolarized and spin-polarized calculations were carried out. For the 1S-1Zr and 1Zr-1Zr zigzag NRs, the total energies of the FM states are lower than those of the nonmagnetic states, indicating that these two kinds of NRs have a magnetic state. However, the other two cases, i.e, armchair and 1S-1S zigzag NRs, are nonmagnetic. First, we focus on the electronic properties of the nonmagnetic armchair and 1S-1S zigzag ZrS$_2$ NRs. The results of the band structures show that all the investigated armchair ZrS$_2$ NRs are indirect-band-gap semiconductors (see Fig. 3(a)). The band gap exhibits interesting odd-even oscillation as increasing the ribbon width (Fig. 3(d)), that is, the even-numbered NRs have relatively larger band gaps than the neighboring odd-numbered NRs. Moreover, the band gap of the even-numbered NRs decreases as a function of the width, while it changes slightly for the odd-numbered NRs. On the other hand, all the 1S-1S zigzag NRs are semiconducting as well but have direct band gaps (Fig. 3(b)). The band gap decreases monotonically as the ribbon width increases (see the inset of Fig. 3(d)). Although the ZrS$_2$ monolayer is an indirect-band-gap semiconductor,\cite{RSCAdv-Li} which makes it less investigated compared with MoS$_2$, the indirect-direct band gap transition can be obtained by cutting it into 1D zigzag NR, with both edges terminated by S atoms .

To investigate the origin of the odd-even oscillation of the band gap, we plot in Fig. 3 (c) the band-decomposed charge densities for armchair ZrS$_2$ NRs with widths $N=6$ and 7. We can see from Fig. 3(a) that for the armchair NRs, the valence band maximum (VBM) locates at the $\Gamma$ point for both the even- and odd-numbered NRs, while the conduction band minimum (CBM) locates at the $X$ points for the odd-numbered NRs and at the position between the $\Gamma$ and $X$ points for the even-numbered ones. The corresponding charge densities for the CBM and VBM are plotted in the upper and lower panels in Fig. 3(c), respectively. We can see that the CBM are mainly determined by the Zr atoms located in the inner part of the NRs and there are much smaller contributions coming from the edge atoms. Therefore, the edge configurations have little impact on the electronic properties of the CBM. In contrast, the VBM are mainly controlled by the S atoms at the edge of the NRs. The structures of the odd-numbered NRs are symmetric with respect to the central line, thus the two nearest-neighboring S atoms at the two edges (which are circled by the green lines) are just opposite to each other, making their distance shortest for the two edge atoms. However, for the even-numbered NRs, the two nearest-neighboring S atoms at the two opposite edges are staggered. Therefore, their distance is enlarged, making the interactions between the two edges relatively smaller. As a result, the VBM is lowered for the even-numbered NRs and their band gaps are enlarged accordingly compared with those of the neighboring odd-numbered NRs.

\begin{figure}[htpb]
\includegraphics[width=0.9\columnwidth]{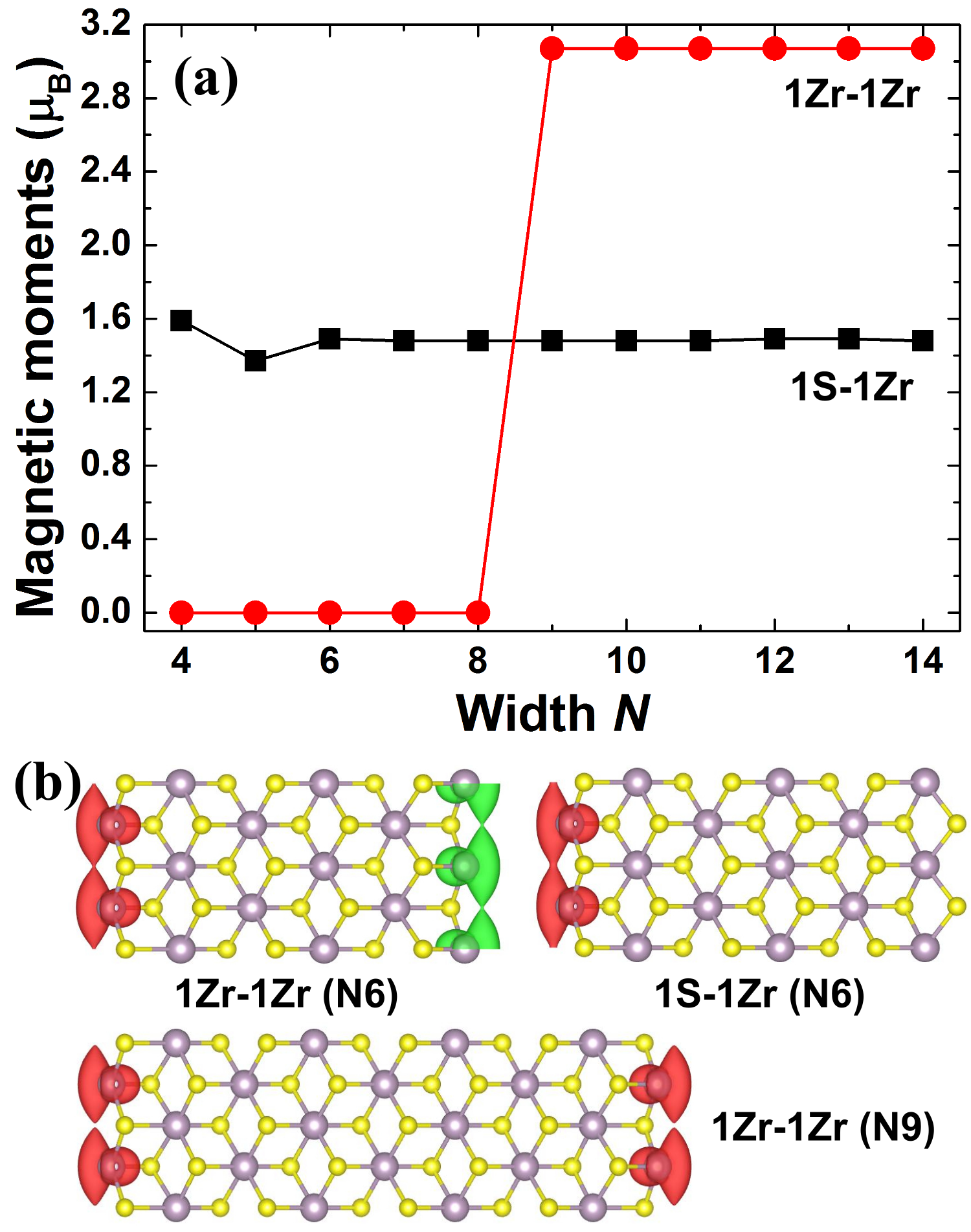}\caption{\label{fig4-magnetic moments}(a) Total magnetic moments of the 1S-1Zr and 1Zr-1Zr zigzag ZrS$_2$ NRs. (b) Spin densities ($\rho$$\uparrow$$-$$\rho$$\downarrow$) for 1Zr-1Zr (N=6), 1S-1Zr (N=6), and 1Zr-1Zr (N=9) NRs.}
\end{figure}

\begin{figure*}[htpb]
\includegraphics[width=1.6\columnwidth]{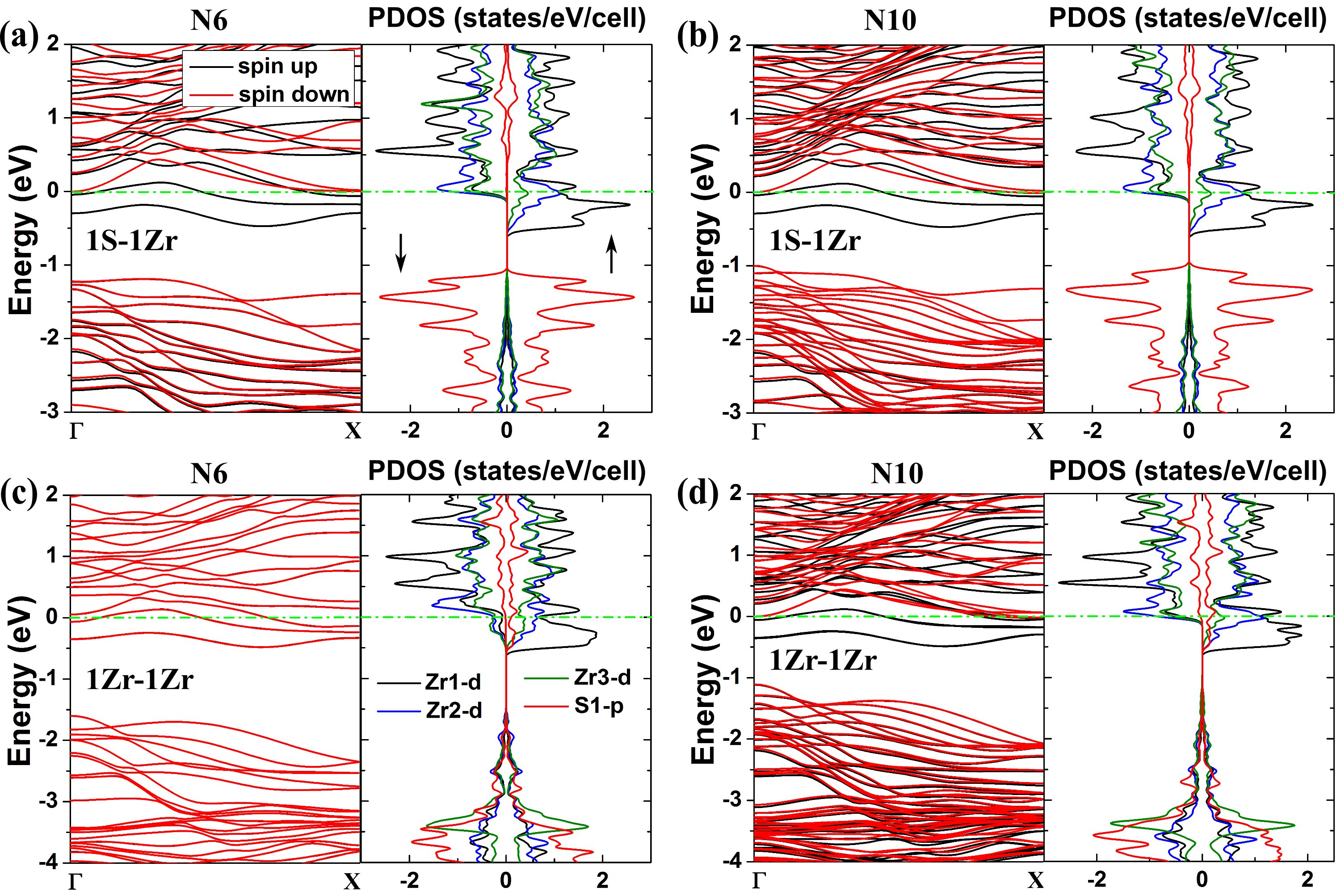}\caption{\label{fig5-band structures}Band structures and partial density of states (PDOS) for the two kinds of magnetic ZrS$_2$ NRs: (a) 1S-1Zr (N=6), (b) 1S-1Zr (N=10), (c) 1Zr-1Zr (N=6), and (d) 1Zr-1Zr (N=10). The black and red lines denote the spin-up and spin-down states, respectively. For the PDOS, the results of four kinds of atoms (denoted in Figs. 1(c) and (d)), that is, the edge S atom (S1), edge Zr atom (Zr1), and two inner Zr atoms near the edge (Zr2 and Zr3) are shown.}
\end{figure*}

As discussed above, the other two kinds of ZrS$_2$ NRs, i.e., 1S-1Zr and 1Zr-1Zr zigzag NRs are magnetic. The results of the spin densities (Fig. 4(b)) demonstrate that the magnetism of both systems mainly originate from the unsaturated edge Zr atoms. This is different from the H-structured zigzag MoS$_2$ NRs, whose magnetic moments concentrate on both the edge Mo and S atoms.\cite{JACS-Chen} For narrow graphene NRs, the spins of the two edges are antiparallel to each other, i.e., the two edges are antiferromagnetically coupled.\cite{PRL-Louie} Since the 1Zr-1Zr ZrS$_2$ NRs have both edges terminated by the Zr atoms, we also check the case when the Zr atoms at the two edges are antiferromagnetically coupled. When the width $N=4-8$, the energies of the AFM states are relatively smaller than those of the FM states and thus the ground states of the NRs are AFM. When the width increases up to 9, the system tends to be FM. The similar property was also observed in zigzag graphene NR, for which the switching from AFM to FM states may occur when the ribbon width is larger than 7 nm, due to the FM inter-edge coupling.\cite{Nature-magda,PRB-Jung} For both the 1S-1Zr and 1Zr-1Zr zigzag NRs, the magnetic moments for the FM states are nearly width-independent, as shown in Fig. 4(a). Since both edges of 1Zr-1Zr NR are terminated by Zr atoms while for 1S-1Zr NR, there is only one edge of Zr atoms, the total magnetic moments of 1Zr-1Zr NRs are twice those of the 1S-1Zr systems.

In Fig. 5, we plot the band structures and partial density of states (PDOS) for the 1S-1Zr and 1Zr-1Zr zigzag NRs with ribbon width $N=6$ and 10. The spin-up and spin-down channels are drawn in the black and red lines, respectively. For the PDOS, the results of four kinds of atoms (denoted in Figs. 1(c) and (d)), including the edge S atom (S1), edge Zr atom (Zr1) and two inner Zr atoms near the edge (Zr2 and Zr3), are shown. For 1S-1Zr ZrS$_2$ NRs (Figs. 5(a) and (b)), the spin-up channels are metallic while the spin-down channels are semiconducting, thus the NRs are 100\% spin-polarized and the systems are FM half-metal. The results of the PDOS show that for S1, the spin-up and spin-down channels are symmetry, so it has no contribution to the magnetic moment. For Zr1, Zr2, and Zr3, the spin-up and spin-down channels are asymmetry. The edge Zr atom (Zr1) contributes most to the magnetic moment and from edge to inner sites, the contribution becomes more and more smaller. Therefore, the magnetic moments mainly originate from the unsaturated Zr atoms located at the edge, which is consistent with the results of spin densities (Fig. 4(b)). For the 1Zr-1Zr ZrS$_2$ NRs, both the spin-up and spin-down states are metallic when $N\leq8$ (Fig. 5(c)), thus the systems are AFM metals. However, the system tends to be FM half-metal when $N$ increases up to 9 and the magnetic moments mainly come from the Zr atoms at the two edges.

Compared with the H-structured MoS$_2$ NRs, the T-structured MX$_2$ NRs are less investigated.  Reyes-Retana $et$ $al.$ reported that the zigzag NiSe$_2$ NRs are metallic and the armchair systems are semiconducting, but both of which are nonmagnetic.\cite{JPCC-Sodi} Our results demonstrate that for the 1S-1Zr and 1Zr-1Zr zigzag ZrS$_2$ NRs, 1D electrical current with completely spin polarization can be realized along the Zr edges of the systems. The intrinsic half-metallicity predicted in the ZrS$_2$ NRs is highly desirable for applications in spintronics.

In conclusion, we have investigated the electronic and magnetic properties of ZrS$_2$ NRs. The armchair ZrS$_2$ NRs are indirect-band-gap semiconductors and the gap exhibits odd-even oscillation as increasing the ribbon width. For the zigzag ZrS$_2$ NRs, the 1S-1S NRs are direct-band-gap semiconductors and the gap decreases with the increase of the width. Both the armchair and 1S-1S zigzag ZrS$_2$ NRs are nonmagnetic. However, the 1S-1Zr and 1Zr-1Zr ($N\geq9$) zigzag NRs are found to be FM half-metal and the magnetism is mainly contributed by the edge Zr atoms. Our results indicate that by tuning the edge configurations, the ZrS$_2$ NRs could exhibit rich electronic and magnetic properties, which is desirable for the future applications in electronics and spintronics.

This work was supported by the National Key Research and Development Program under Contract No. 2016YFA0300404, National Natural Science Foundation of China under Contract Nos. 11404340, 11674326, and U1232139, the Key Research Program of Frontier Sciences of CAS (QYZDB-SSW-SLH015) and Hefei Science Center of CAS (2016HSC-IU011), the Anhui Provincial Natural Science Foundation under Contract No. 1708085QA18. The calculation was partially performed at the Center for Computational Science, CASHIPS.

\end{document}